\documentclass[prb,twocolumn,superscriptaddress,amsmath,amssymb]{revtex4}

\newcommand{\fatr}{\mathbf{r}}

\usepackage{array}
\usepackage{graphicx}
\usepackage{color}
\usepackage{dcolumn}
\usepackage{bm}

\def\beq{\begin{equation}}
\def\eeq{\end{equation}}
\def\bea{\begin{eqnarray}}
\def\eea{\end{eqnarray}}

\def\fatr{{\bf r}}
\def\fatx{{\bf x}}

\begin{document}
\title{Force correcting atom centered potentials for generalized gradient approximated density functional theory: 
Approaching hybrid functional accuracy for geometries and harmonic frequencies in small chlorofluorocarbons}
\author{O.~Anatole von Lilienfeld}
\email{anatole@alcf.anl.gov}
\affiliation{Argonne Leadership Computing Facility, Argonne National Laboratory, Argonne, Illinois 60439, USA}
\affiliation{Department of Chemistry, University of Basel, Klingelbergstr. 80, 4056 Basel, Switzerland\footnote{as of summer 2013}}
\date{\today}

\begin{abstract}
Generalized gradient approximated (GGA) density functional theory (DFT) typically overestimates polarizability and bond-lengths, and underestimates force constants of covalent bonds.
To overcome this problem we show that one can use empirical force correcting atom centered potentials (FCACPs), parameterized for every nuclear species. 
Parameters are obtained through minimization of a penalty functional that explicitly encodes hybrid DFT forces and static polarizabilities of reference molecules.  
For hydrogen, fluorine, chlorine, and carbon the respective reference molecules consist of H$_2$, F$_2$, Cl$_2$, and CH$_4$. 
The transferability of this approach is assessed for harmonic frequencies in a small set of chlorofluorocarbon molecules.
Numerical evidence, gathered for CF$_4$, CCl$_4$, CCl$_3$F, CCl$_2$F$_2$, CClF$_3$, ClF, HF, HCl, CFH$_3$, CF$_2$H$_2$, CF$_3$H, CHCl$_3$, CH$_2$Cl$_2$, CH$_3$Cl indicates that the GGA+FCACP level of theory yields harmonic frequencies 
that are significantly more consistent with hybrid DFT values, as well as slightly reduced molecular polarizability. 
\end{abstract}

\maketitle

\section{Introduction}
The rigorous analysis of theoretical predictions and experimental measurements for molecular properties is one of the noblest tasks in physical chemistry.
The fundamental importance of this line of research can hardly be overstated.
To paraphrase M.~Quack from one of his physical chemistry lectures at ETH Z\"urich, 
{\em we only understand molecules when we are able to predict their properties to a degree considered quantitative}.
Philosophically it is compelling to note that this definition distinguishes predictive power as a sufficient criterion for understanding.
Conversely, qualitative predictions would indicate an only incomplete understanding.  
We also note that this definition qualifies ``understanding'' in terms of an arbitrarily chosen accuracy 
criterion for what, in some context of usefulness, is deemed sufficiently quantitative. 

Over the last decades, members of the laboratory of M.~Quack have made major contributions to enable and carry out 
quantitative predictions for small and isolated molecules, using 
high-resolution molecular spectroscopy and highly accurate quantum chemical approaches to address
fundamental questions such as parity violation~\cite{QuacksParity,QuackPerspectiveParityViolation2011}. 
When it comes to condensed phase spectroscopy, however, we ordinarily have to rely on less accurate
generalized gradient approximations (GGA) within density functional theory (DFT)~\cite{HK,KS} 
based molecular dynamics (AIMD)~\cite{AIMD_PNAS_TUCKERMAN2005}.
See for example the recent contributions by M.~P.~Gaigeot and others concerning infra red spectra of small peptides in liquid 
water~\onlinecite{GaigeotVuilleumierInfraRed2007,GaigeotAlanineDipeptide2010,GaigeotFloppyPeptide2010}, as well as references therein.
Vibrational spectra of surfaces can also be calculated using DFT~\cite{IRiceSurface_Galli2012}.
And even for small molecules adsorbed on surfaces, the computational pediction of infra red spectra has already been evinced, 
e.g.~water adsorbed on Ni (111) and (211)~\cite{luigi-niwater,tatiana-niwater-frequencies},
or organic molecules adsorbed on Si (111)~\cite{VibrationalAlkylAdsorbedSi_Galli2012}.

Unfortunately, GGAs are not always sufficiently accurate when it comes to vibrational properties,
and admixture of Hartree-Fock exchange in the form of hybrid functionals might be necessary to 
yield significantly more accurate predictions~\cite{ChemistsGuidetoDFT}, 
as also confirmed for the vibrational spectra of CCl$_3$F~\cite{anatole-pccp2007}.
For plane-wave basis calculations of condensed phase systems under periodic boundary conditions, 
however, the calculation of the exchange term is computationally dramatically more expensive than pure GGA, typically by an order of magnitude. 
One strategy to overcome this challenge consists of implementing sophisticated and highly parallelized 
software that can efficiently exploit high-performance compute hardware such as IBM's BlueGene~\cite{INCITEprogram}, 
with tens of thousands of compute nodes, e.g.~{\tt CPMD}~\cite{cpmd3.15,VeryLargeScaleCurioni_2010} 
or {\tt Qbox}~\cite{Qbox_Gygi2008,HybridDFTQbox_Gygi2010}.
While hybrid functionals are implemented and not impossible to use for condensed systems, 
as demonstrated in studies of liquid water~\cite{Hybridwater}, and its IR spectrum~\cite{HybridIRwater_GalliGygi2011},
access to the substantial CPU resources required is as restricted as is the number of possible systems that can be tackled.

At the hybrid DFT level of theory, comparative studies of a multitude of condensed systems are therefore prohibitive, 
let alone any exploration attempts in chemical compound space for bio or materials design from first principles~\cite{CCSanatole2012}.
Alternatively, one can also attempt to improve the GGA's accuracy without increasing its computational complexity.  

In this article, we will discuss and investigate a force correcting atom centered potentials (FCACPs) approach. 
The goal is to augment GGA calculations to approach hybrid DFT accuracy at negligible additional computational cost. 
We will show that for all the molecules studied, GGA+FCACP consistently yields geometries and harmonic frequencies 
with near hybrid DFT accuracy. 

\section{Method}
\subsection{Background}
Within the Born-Oppenheimer approximation, Hellmann-Feynman's theorem clearly states that the 
forces on atoms are due to the electrostatic field exerted by the ground-state electron density~\cite{HF}. 
Since we ignore the explicit form of the exact exchange-correlation potential~\cite{KS} one can argue
that it is reasonable to empirically manipulate the density with the objective to yield forces that come as close as possible to known reference values.
An appealing way to induce such changes into the electron density consists of adding atom centered non-local potentials 
that become negligibly small at the site of the nucleus, not to affect the core electrons, 
but are of sufficient magnitude in the interatomic region where the covalent bonding occurs. 
These correcting potentials are similar to pseudopotentials (PPs), and, while not necessarily so, can even assume their functional form.
PPs, or effective core potentials, replace the explicit treatment of the core 
electrons~\cite{Hellmann1,Hellmann2,KleinmansPP,WeeksPP,bhs,ChristiansensPP,DolgsECPs}, thereby reducing
(i) the number of orbitals to be dealt with, and (ii) dramatically accelerating basis-set convergence in plane-wave basis sets.

The idea to adapt PPs to account also for other properties, i.e.~going beyond the mere purpose of modeling the core electrons' potential, is not new.
It has successfully been deployed for relativistic effects~\cite{RelativisticPPs_bs1982,sgpsp,RelativisticPPreview2012}, 
self-interaction corrections,~\cite{SICPP-Vogl,SICPP-Pollmann}
modeling exact-exchange electron densities~\cite{anatole-jcp2005}, 
atomization energies and geometries of Al-clusters~\cite{AtomizationEnergiesAlPPs_Truhlar2005jctc},
and minimizing quantum mechanical/molecular mechanical boundary 
errors~\cite{Christiansen-JCP-2002,anatole-jcp2005,khong-cappingpotentials,QMMMcapping_schiffmannSebastiani2011,QMMMcapping_Sebastiani2011},
widening the band gap~\cite{Christensen1984,vandewalle2007,anatole-prb2008},
and introducing van-der-Waals 
interactions~\cite{anatole-prl2004,anatole-prb2005,anatole-prb2007,DiLabioDCACP2008,DiLabioDCACPapplication2008,DiLabioDCACP2012}. 
Various recent applications demonstrate the success of the latter, including binding of ellipticine to DNA and 
other biomolecular interactions~\cite{anatole-jpcb2007}, and the accurate description of molecular crystals~\cite{enrico-jctc2007,DCACP4molcrystals} 
Furthermore, one can interpolate pseudopotentials, and perform self-consistent field calculations as a function 
of order parameter, $0 \le \lambda \le 1$,  effectively corresponding to fractional nuclear charges~\cite{anatole-prl2005,anatole-jcp2006-2,anatole-jcp2009-2}. 
The effect of such ``alchemical'' variations on hydrogen-bonded dimers was investigated in combination with 
atom centered van der Waals correction~\cite{anatole-jctc2007}. 
Alchemical changes, and corresponding Hellmann-Feynman derivatives~\cite{anatole-jcp2009-2}, are commonly 
used for two, often related, purposes: Either for the evaluation of free energy differences
between different compounds, e.g.~using thermodynamic integration~\cite{TI}, $\Delta F = \int d\lambda \; \langle \partial E/\partial \lambda \rangle$,
e.g.~see Ref.~\onlinecite{anatole-jcp2009};
or for obtaining gradients that quantify a system's response to a variation in chemical composition~\cite{AlchemicalDerivativeBinaryMetalCluster_WeigendSchrodtAhlrichs2004,anatole-prl2005,RCD_Yang2006,anatole-jcp2006-2,anatole-jcp2007,CatalystSheppard2010}.

\subsection{Optimization}
The goal of this study is to explore if yet another property can be optimized through manipulation of PP parameter space, 
namely the force in the covalent bond. 
While at first GGAs can be considered sufficiently accurate, their interatomic distances, polarizabilities, and vibrational frequencies
are typically considerably off when compared to more accurate methods, such as hybrid DFT results~\cite{ChemistsGuidetoDFT}. 
While the choice of reference method and geometry is somewhat {\em ad hoc}, for this study  
the non-empirical hybrid functional PBE0 has been selected~\cite{PBE0,PBE01,PBE} without any loss of fundamental generality: 
Post-Hartree-Fock methods such as Coupled Cluster, or quantum Monte Carlo, could have been chosen just as well.
Homo-diatomics have been used as reference systems, i.e. H$_2$, F$_2$, and Cl$_2$ 
for parameterizing the hydrogen, fluorine, and chlorine atom.
Since the smallest stable molecule consisting exclusively of carbon is the exceedingly large C$_{20}$ buckyball, 
methane has been chosen as a reference system instead. 
For the hydrogen atoms the previously optimized FCACPs are used without any further changes.

For the optimization a unitless penalty functional, $\mathcal{P}$, is defined that allows for multi-objective 
optimization of $N$ normalized properties in PP parameters $\fatx$,
\bea
\min_{\fatx} \mathcal{P}(\fatx) & = & \min_{\fatx} \frac{1}{N} \sum_i^N \frac{|P_i^{\rm ins}(\fatx)-P_i^{\rm ref}|}{|P_i^{\rm ini}-P_i^{\rm ref}|},
\label{eq:Penalty}
\eea
where $P_i^{\rm ins}, P_i^{\rm ini}, P_i^{\rm ref}$ are the instantaneous, initial, and reference values of property $i$, respectively.
For this study Eq.~(\ref{eq:Penalty}) has been chosen to minimize the deviation from two reference properties calculated with the hybrid functional 
PBE0~\cite{PBE0,PBE01,PBE}, (i) the 2-norm of all ionic forces of the system in the
reference geometry, $|F|$, and (ii) the trace of the static polarizability tensor, Tr($\alpha$).
The choice of the latter is based on the fact that the variational principle also holds in terms of Pearsson's maximum hardness 
principle~\cite{HSABPearson,MaxHardnessPearson,MaxHardnessParr}, an alternative to the potential energy 
whose origin is arbitrary in plane-wave PP based calculations. 

The minimization in PP parameter space could have been carried out with steepest descent or conjugate
gradient algorithms using property derivatives calculated through linear response~\cite{apdsmp}, 
as proposed in Ref.~\onlinecite{anatole-jcp2005}. 
To facilitate the implementation, however, a gradient-free optimizer has been used for this study,
namely Nelder-Mead's simplex optimization method~\cite{SimplexNelderMead}.
The FCACP assumes the form of the highest empty angular momentum channel $l$ for each atom type 
in the form of a Goedecker-Hutter non-local PP~\cite{SG}, 
in close analogy to the DCACP parameterization ~\cite{anatole-prb2007},
Specifically, $l = s = 1$ for H, and $l = d = 3$ for C, F, and Cl.
As such, this defines a 2-dimensional parameter space of atom centered potential parameters, $\fatx = (x_1,x_2)$, for each atom type.
Here, $x_1 = r_l$, the Gaussian width in the non-local projector $p^{lm}$,~\cite{SG} 
\bea
\langle \fatr | p_{lm} \rangle & = & N_l Y_{lm}(\hat{r}) r^l e^{-\frac{r^2}{2 r^2_l}},
\eea
where $N_l$ and $Y_{lm}(\hat{r})$ correspond to the normalization constant and spherical harmonics, and $r$ is the radial distance from the atom.
While this projector is centered on the atom, the multiplication of the polynomial $r^l$ term scales it down to zero at the position of 
the nucleus. $x_1$ thus tunes the location of the projector's maximum at a distance from the atom. 
The second parameter, $x_2 = h_l$, scales the magnitude of the entire nonlocal pseudopotential contribution from the correcting channel, 
\bea
V_{\rm FCACP}(\fatr,\fatr') & = & \sum_{lm} \langle \fatr | p_{lm} \rangle h_l \langle p_{lm} | \fatr' \rangle.
\eea

\subsection{Computational details}
All the DFT reference and optimization calculations have been carried out with the {\tt CPMD} PP plane-wave basis set program~\cite{cpmd3.13.2}.
Polarizabilities are obtained through the linear response tools implemented in {\tt CPMD}~\cite{apdsmp}, 
and  harmonic frequencies in {\tt CPMD} are calculated from Hessians obtained via finite differences. 
PBE Goedecker-Hutter PPs have been used for all the calculations~\cite{SG,KrackPP}.
While the use of GGA PPs within hybrid DFT, or even within GGA + atom centered corrections, is not unconventional, 
we note that eventually the PPs should be reparameterized to be entirely consistent with their density functional~\cite{MeaningfulDFTMaterials2005}.
All calculations involved isolated boundary conditions with the Poisson solver by Martyna and Tuckerman~\cite{martyna-tuckerman}, 
a plane-wave cutoff of 200 Ry, a unit-cell of 15$\times$15$\times$15 {\AA}$^3$, and were carried out on
Argonne Leadership Computing Facility's IBM BlueGene/P machine.
At this point the reader is cautioned that severe finite size effects in the plane-wave calculations can 
lead to significant distortions of polarizabilities and frequencies. 
This is of little concern for this study, however, since we only deal with relative changes in these properties, and 
the finite size effects can be assumed to cancel when comparing results from different functionals. 

\section{Results and discussion}
\subsection{Optimization for reference molecules}
\begin{table}
\caption{
Converged PBE+FCACP parameters for hydrogen, fluorine, chlorine and carbon atoms, respectively optimized for 
reference systems H$_2$, F$_2$, Cl$_2$, and CH$_4$ in PBE0 geometry.
$|F|$(P+F) denotes the residual Euclidean norm of PBE+FCACP forces in PBE0 relaxed geometry in Hartree/{\AA}. 
$\Delta$ shows the deviation of trace of static polarizability from PBE0, 
for PBE (P) and PBE+FCACP (P+F), respectively.
[Bohr$^3$] are shown as well. 
$l$ = 1, 3, 3, 3, for H, F, Cl, and C, respectively.
}
\label{tab:params}
\begin{tabular}{l|r@{.}l|r@{.}l|r@{.}l|r@{.}l|r@{.}l} \hline
Atom&  \multicolumn{2}{c}{$r_l$[a.u.]} &  \multicolumn{2}{c}{$h_l$[a.u.]}    &  \multicolumn{2}{c}{$|F|$(P+F)} & \multicolumn{2}{c}{$\Delta$(P+F)} & \multicolumn{2}{c}{$\Delta$(P)} \\\hline
H   &  0&9871& -0&004129 & 3&8$\times 10^{-7}$   & 1&01 & 1&15 \\
F   &  1&3343& -0&014713 & 5&5$\times 10^{-7}$   & 1&85 & 2&02\\
Cl  &  1&3199& -0&006414 & 3&9$\times 10^{-7}$   & 5&94 & 6&12\\
C   &  0&8081& -0&035297 & 3&8$\times 10^{-4}$   & 3&06 & 3&43\\ \hline
\end{tabular}
\end{table}

\begin{table}[ht!]
\caption{
Results for four molecules used as reference systems for optimization. 
Dipole moments, polarizabilities and harmonic frequencies with PBE, PBE0, and PBE+FCACP. 
PBE+FCACP results for PBE0 geometries.
Traces of polarizabilities $|\alpha|$ in atomic units. 
Harmonic frequencies $\omega$ [cm$^{-1}$] in descending order of energy. 
If degenerate the averaged value is reported, $d$ and $t$ denoting doublet or triplet.
Experimental wavenumbers $\nu$, all in [cm$^{-1}]$ and from Ref.~\cite{NISTvibrations}, are given as a footnote for orientation.
}
\label{tab:results}
\begin{tabular}{lc|r@{.}l|r@{.}lr@{.}lr@{.}lr@{.}l} \hline
Molecule &  method &   \multicolumn{2}{c}{$|\alpha|$} & \multicolumn{2}{c}{$\omega_1$} & \multicolumn{2}{c}{$\omega_2$} & \multicolumn{2}{c}{$\omega_3$} & \multicolumn{2}{c}{$\omega_4$} \\ \hline
H$_2$\footnote{4401.2; $^b$ 916.9; $^c$ 559.7; $^{\rm d}$3019.0, 2917, 1534.0, 1306.0}   &   PBE    & 6&84 & 4306&6       & \multicolumn{2}{c}{}      & \multicolumn{2}{c}{}      & \multicolumn{2}{c}{}      \\ 
&PBE+FCACP & 6&66 & 4386&2       & \multicolumn{2}{c}{} & \multicolumn{2}{c}{}& \multicolumn{2}{c}{}\\ 
&   PBE0   & 4&40 & 4402&7       & \multicolumn{2}{c}{}& \multicolumn{2}{c}{}& \multicolumn{2}{c}{}\\\hline 
F$_2$$^b$   &   PBE    &12&99 & 1087&2   & \multicolumn{2}{c}{}& \multicolumn{2}{c}{}& \multicolumn{2}{c}{}\\ 
&PBE+FCACP& 12&49 & 1155&2   & \multicolumn{2}{c}{}& \multicolumn{2}{c}{}& \multicolumn{2}{c}{} \\ 
&   PBE0   & 6&80 & 1139&2   & \multicolumn{2}{c}{}& \multicolumn{2}{c}{}& \multicolumn{2}{c}{}\\\hline 
Cl$_2$$^c$  &   PBE    &41&92 & 538&6   & \multicolumn{2}{c}{}& \multicolumn{2}{c}{}& \multicolumn{2}{c}{} \\
&PBE+FCACP& 41&47 & 561&2   & \multicolumn{2}{c}{}& \multicolumn{2}{c}{}& \multicolumn{2}{c}{}\\
&   PBE0   &24&56 & 578&1   & \multicolumn{2}{c}{}& \multicolumn{2}{c}{}& \multicolumn{2}{c}{} \\\hline 
CH$_4$$^{\rm d}$&   PBE    &16&87 & 3075&56$^t$ & 2951&3 & 1494&3$^d$ & 1277&5$^t$ \\ 
&PBE+FCACP& 16&50 & 3142&0$^t$ & 3017&9 & 1514&7$^d$ & 1287&8$^t$ \\ 
&   PBE0   &13&44 & 3157&2$^t$ & 3029&8 & 1539&7$^d$ & 1316&9$^t$ \\ \hline \hline 
\end{tabular}
\end{table}
Optimized FCACP parameters are reported in Table~\ref{tab:params} for hydrogen, fluorine, chlorine, and carbon.
The positioning parameter, $r_l$, converges to a length scale of 0.8 to 1.4 Bohr for all atoms. 
For comparison, the actual PP parameters in the non-local PP $s$-channel (not existent for the hydrogen's PP) 
range from 0.2 to 0.3 Bohr for C, F, and Cl.
By contrast, the DCACP values for the $f$-channel range from 1.8 to 3.6 Bohr~\cite{anatole-prb2007,DCACPsulfur-jctc2009}.
It is thus clear that the FCACP corrections acts in a more intermediate mid-range around the atoms. 
This range of action is not surprising, it is roughly the distance from atom to covalent bonding.
Similar comparisons can be made for the magnitude of the correction, $h_l$: 
The $s$-channel of the C, F, and Cl PPs ranges from 9.6 to 23.7 a.u. in magnitude, 
the DCACP's $f$-channel is typically of only $\sim$ -10$^{-4}$ a.u.~\cite{anatole-prb2007,DCACPsulfur-jctc2009}. 
The FCACP's $h_l$ is in between at $\sim$ -10$^{-2}$ to -10$^{-3}$ a.u.
It is not surprising that the FCACPs are one to two orders of magnitude larger than the DCACPs  
since the absolute errors in covalent forces are significantly larger than the GGA's error to account for van der Waals forces.
As such, FCACPs correspond to an attractive potential that counteracts the effect of an overly delocalized GGA electron density 
on interatomic covalent forces. 

Table~\ref{tab:params} also enlists the residual forces and polarizabilities in the corresponding reference molecules after convergence of the FCACPs. 
We note that on the one hand the norm of the forces can be quenched to virtually correspond to the reference method's force 
(zero in this study) within numerical precision, with the slight exception of carbon where convergence sets in already at $\sim$10$^{-4}$ Hartree/{\AA}. 
The uncorrected PBE forces in PBE0 geometry typically amount to 10$^{-2}$ to 10$^{-3}$ Hartree/{\AA}.
On the other hand, however, the static polarizability's trace does not improve as much, the PBE value is usually reduce by no more than 10\%.
Additional optimization test runs suggest that removal of the force from the objective penalty functional in 
Eq.~(\ref{eq:Penalty}) would yield perfect reproduction of the reference polarizability, 
at the expense, however, that the resulting forces worsen considerably. 
As such, polarizability appears to be crucial as a constraint for the main objective of having correct forces in the covalent bond. 
This finding underscores the importance of polarizability, as already widely researched and discussed 
in terms of the maximal hardness principle and the hard-soft acid-base principles by Pearson, 
and Parr and Chattaraj~\cite{HSABPearson,MaxHardnessPearson,MaxHardnessParr}.

\subsection{Frequencies for reference molecules}
Table~\ref{tab:results} reports final polarizability traces, as well as harmonic frequencies for the 
four reference molecules used for optimization, i.e.~H$_2$, F$_2$, Cl$_2$, and CH$_4$.
We note that even though frequencies were not explicitly encoded in the penalty minimization,
for all cases the PBE+FCACP frequencies are significantly closer to PBE0.
For all frequencies except the one for F$_2$, PBE+FCACP approaches the PBE0 value from below.
Already this suggests that F containing molecules may be harder to deal with, {\em vide infra}.

\subsection{Transferability to other molecules}

\begin{table*}
\caption{
Results for test molecules for which the FCACPs have been used without any further changes.
Dipole moments, polarizabilities and harmonic frequencies with PBE, PBE0, and PBE+FCACP.
PBE+FCACP results for PBE0 geometries.
Total dipole moments $\mu$ and averaged traces of polarizability $|\alpha|$ in atomic units. 
Harmonic frequencies $\omega$ [cm$^{-1}$] in descending order of energy. 
If degenerate the averaged value is reported, $d$ and $t$ denoting doublet or triplet.
Experimental wavenumbers $\nu$, all in [cm$^{-1}]$ and from Ref.~\onlinecite{NISTvibrations}, are given as a footnote for orientation.
The mean absolute error from PBE0 using either PBE or PBE+FCACP is shown in the last two lines, respectively. 
}
\label{tab:testresults}
\begin{tabular}{lc|r@{.}l|r@{.}l|r@{.}lr@{.}lr@{.}lr@{.}lr@{.}lr@{.}lr@{.}lr@{.}lr@{.}l} \hline
Molecule &  method &  \multicolumn{2}{c}{$\mu$}  & \multicolumn{2}{c}{$|\alpha|$} & \multicolumn{2}{c}{$\omega_1$} & \multicolumn{2}{c}{$\omega_2$} & \multicolumn{2}{c}{$\omega_3$} & \multicolumn{2}{c}{$\omega_4$} & \multicolumn{2}{c}{$\omega_5$} & \multicolumn{2}{c}{$\omega_6$} & \multicolumn{2}{c}{$\omega_7$} & \multicolumn{2}{c}{$\omega_8$} & \multicolumn{2}{c}{$\omega_9$}\\ \hline
HF
\footnote{4138.4; $^b$2990.9;$^c$738.5; $^{\rm d}$776.0, 459.0, 314.0, 217.0; $^e$1280.0, 909.0, 631.0, 453.0; $^f$1085.0, 847.0, 535.0, 394.0, 350.0, 241.0; $^g$3034.1, 1219.7, 774.0, 680.0, 366.0, 260.0; $^h$3006.0, 2930.0, 1467.0, 1464.0, 1182.0, 1049.0; $^i$1212.0, 1105.0, 781.0, 563.0, 476.0, 350.0; $^j$3041.8, 2966.2, 1454.6, 1354.9, 1015.0, 732.1; $^k$3036.0, 1372.0, 1152.0, 1117.0, 700.0, 507.0; $^l$1159.0, 1101.0, 902.0, 667.0, 458.0, 446.0, 437.0, 322.0, 262.0; $^m$3040.0, 2999.0, 1467.0, 1268.0, 1153.0, 898.0, 758.0, 717.0, 282.0, $^n$3014.0, 2948.0, 1508.0, 1435.0, 1262.0, 1178.3, 1111.2, 1090.1, 528.5} 
&   PBE    & 0&71 & 6&02  & 3932&2  & \multicolumn{2}{c}{}&\multicolumn{2}{c}{}&\multicolumn{2}{c}{}&\multicolumn{2}{c}{}&\multicolumn{2}{c}{}&\multicolumn{2}{c}{}&\multicolumn{2}{c}{}&\multicolumn{2}{c}{}\\  
&PBE+FCACP & 0&68 & 5&99  & 4233&2  & \multicolumn{2}{c}{}&\multicolumn{2}{c}{}&\multicolumn{2}{c}{}&\multicolumn{2}{c}{}&\multicolumn{2}{c}{}&\multicolumn{2}{c}{}&\multicolumn{2}{c}{}&\multicolumn{2}{c}{}\\  
&   PBE0   & 0&72 & 4&69  & 4231&9  & \multicolumn{2}{c}{}&\multicolumn{2}{c}{}&\multicolumn{2}{c}{}&\multicolumn{2}{c}{}&\multicolumn{2}{c}{}&\multicolumn{2}{c}{}&\multicolumn{2}{c}{}&\multicolumn{2}{c}{}\\ \hline 
HCl$^b$ &   PBE    & 0&43 &17&58  & 2900&4 & \multicolumn{2}{c}{}&\multicolumn{2}{c}{}&\multicolumn{2}{c}{}&\multicolumn{2}{c}{}&\multicolumn{2}{c}{}&\multicolumn{2}{c}{}&\multicolumn{2}{c}{}&\multicolumn{2}{c}{}\\ 
&PBE+FCACP & 0&41 &17&52  & 2989&0 & \multicolumn{2}{c}{}&\multicolumn{2}{c}{}&\multicolumn{2}{c}{}&\multicolumn{2}{c}{}&\multicolumn{2}{c}{}&\multicolumn{2}{c}{}&\multicolumn{2}{c}{}&\multicolumn{2}{c}{}\\ 
&   PBE0   & 0&44 &14&17  & 2999&5 & \multicolumn{2}{c}{}&\multicolumn{2}{c}{}&\multicolumn{2}{c}{}&\multicolumn{2}{c}{}&\multicolumn{2}{c}{}&\multicolumn{2}{c}{}&\multicolumn{2}{c}{}&\multicolumn{2}{c}{}\\ \hline
ClF$^c$ &   PBE    & 0&33 &18&64  &   791&0 & \multicolumn{2}{c}{}& \multicolumn{2}{c}{}& \multicolumn{2}{c}{} & \multicolumn{2}{c}{}& \multicolumn{2}{c}{}& \multicolumn{2}{c}{} & \multicolumn{2}{c}{}& \multicolumn{2}{c}{}\\ 
&PBE+FCACP & 0&31 &18&47  &   840&6 & \multicolumn{2}{c}{}& \multicolumn{2}{c}{}& \multicolumn{2}{c}{} & \multicolumn{2}{c}{}& \multicolumn{2}{c}{}& \multicolumn{2}{c}{} & \multicolumn{2}{c}{}& \multicolumn{2}{c}{}\\ 
&   PBE0   & 0&33 &14&75  &   852&7 & \multicolumn{2}{c}{}& \multicolumn{2}{c}{}& \multicolumn{2}{c}{} & \multicolumn{2}{c}{}& \multicolumn{2}{c}{}& \multicolumn{2}{c}{} & \multicolumn{2}{c}{}& \multicolumn{2}{c}{}\\\hline \hline 
CCl$_4$$^{\rm d}$&   PBE    & 0&02 &70&40  & 729&1$^t$ & 440&1 & 300&4$^t$ & 199&6$^d$ & \multicolumn{2}{c}{}&\multicolumn{2}{c}{}&\multicolumn{2}{c}{}&\multicolumn{2}{c}{}&\multicolumn{2}{c}{}\\ 
&PBE+FCACP & 0&02 &69&85  & 745&6$^t$ & 445&4 & 306&8$^t$ & 204&5$^d$ & \multicolumn{2}{c}{}&\multicolumn{2}{c}{}&\multicolumn{2}{c}{}&\multicolumn{2}{c}{}&\multicolumn{2}{c}{}\\ 
&   PBE0   & 0&02 &55&46  & 799&4$^t$ & 467&6 & 315&6$^t$ & 212&1$^d$ & \multicolumn{2}{c}{}&\multicolumn{2}{c}{}&\multicolumn{2}{c}{}&\multicolumn{2}{c}{}&\multicolumn{2}{c}{}\\ \hline 
CF$_4$$^e$  &   PBE    & 0&01  &20&38 &1202&8$^t$ & 906&7 & 687&0$^t$ & 566&9$^d$ & \multicolumn{2}{c}{}&\multicolumn{2}{c}{}&\multicolumn{2}{c}{}&\multicolumn{2}{c}{}&\multicolumn{2}{c}{}\\  
&PBE+FCACP & 0&01  &20&04 &1230&1$^t$ & 862&7 & 639&5$^t$ & 496&4$^d$ & \multicolumn{2}{c}{}&\multicolumn{2}{c}{}&\multicolumn{2}{c}{}&\multicolumn{2}{c}{}&\multicolumn{2}{c}{}\\  
&   PBE0   & 0&01 &15&91  &1291&6$^t$ & 930&9 & 677&9$^t$ & 539&6$^d$ & \multicolumn{2}{c}{}&\multicolumn{2}{c}{}&\multicolumn{2}{c}{}&\multicolumn{2}{c}{}&\multicolumn{2}{c}{}\\ \hline\hline 
CCl$_3$F$^f$&   PBE    & 0&19 &57&83 & 1046&2   & 797&5$^d$ &  537&1 & 458&1$^d$ & 354&8  & 250&0$^d$ &\multicolumn{2}{c}{}&\multicolumn{2}{c}{}&\multicolumn{2}{c}{}\\  
&PBE+FCACP & 0&19 &57&34 & 1053&4   & 826&8$^d$ &  541&4 & 464&6$^d$ & 359&8  & 257&3$^d$ &\multicolumn{2}{c}{}&\multicolumn{2}{c}{}&\multicolumn{2}{c}{}\\  
&   PBE0   & 0&19 &45&47 & 1124&5   & 862&1$^d$ &  563&6 & 463&4$^d$ & 369&5  & 262&2$^d$&\multicolumn{2}{c}{}&\multicolumn{2}{c}{}&\multicolumn{2}{c}{}\\  \hline 
CHCl$_3$$^g$&   PBE    & 0&40 &57&11 & 3066&3  &1187&4$^d$ & 722&3$^d$ & 657&2  & 352&7    & 245&4$^d$&\multicolumn{2}{c}{}&\multicolumn{2}{c}{}&\multicolumn{2}{c}{}\\ 
&PBE+FCACP & 0&37 &56&66 & 3138&7  &1200&7$^d$ & 734&9$^d$ & 670&3  & 360&5    & 250&4$^d$&\multicolumn{2}{c}{}&\multicolumn{2}{c}{}&\multicolumn{2}{c}{}\\ 
&   PBE0   & 0&42 &45&02 & 3159&9  &1252&5$^d$ & 788&9$^d$ & 689&7  & 371&8    & 257&9$^d$ &\multicolumn{2}{c}{}&\multicolumn{2}{c}{}&\multicolumn{2}{c}{}\\ \hline 
CFH$_3$$^h$ &   PBE    & 0&73 &17&48 & 3055&5$^d$ & 2946&1 & 1531&8$^d$ & 1481&9 & 1291&9$^d$ & 1065&9 &\multicolumn{2}{c}{}&\multicolumn{2}{c}{}&\multicolumn{2}{c}{}\\ 
& PBE+FCACP& 0&70 &17&14 & 3133&8$^d$ & 3023&9 & 1556&8$^d$ & 1494&7 & 1331&5$^d$ & 1062&8 &\multicolumn{2}{c}{}&\multicolumn{2}{c}{}&\multicolumn{2}{c}{}\\ 
&   PBE0   & 0&73 &13&74 & 3143&8$^d$ & 3034&8 & 1553&4$^d$ & 1522&6 & 1318&1$^d$ & 1126&8 &\multicolumn{2}{c}{}&\multicolumn{2}{c}{}&\multicolumn{2}{c}{}\\ \hline
CClF$_3$$^i$&   PBE    & 0&19 &32&60 &1147&7$^d$& 1051&5 &  799&0& 639&3$^d$ & 483&3 & 446&8$^d$ &\multicolumn{2}{c}{}&\multicolumn{2}{c}{}&\multicolumn{2}{c}{}\\ 
&PBE+FCACP & 0&18 &32&17 &1169&0$^d$& 1070&8 &  770&3& 568&0$^d$ & 455&5 & 355&1$^d$ &\multicolumn{2}{c}{}&\multicolumn{2}{c}{}&\multicolumn{2}{c}{}\\ 
&   PBE0   & 0&19 &25&54 &1231&3$^d$& 1114&7 &  800&3& 584&3$^d$ & 480&5 & 365&0$^d$ &\multicolumn{2}{c}{}&\multicolumn{2}{c}{}&\multicolumn{2}{c}{}\\ \hline 
CH$_3$Cl$^j$&   PBE    & 0&74 &29&67 &3084&8$^d$& 2983&2 & 1419&2$^d$& 1326&3    & 999&1$^d$ & 716&7 &\multicolumn{2}{c}{}&\multicolumn{2}{c}{}&\multicolumn{2}{c}{}\\ 
&PBE+FCACP & 0&71 &29&33 &3154&2$^d$& 3053&0 & 1434&2$^d$& 1336&7    &1014&4$^d$ & 720&7 &\multicolumn{2}{c}{}&\multicolumn{2}{c}{}&\multicolumn{2}{c}{}\\ 
&   PBE0   & 0&75 &23&65 &3174&2$^d$& 3069&0 & 1468&3$^d$& 1376&7    &1038&7$^d$ & 757&3 &\multicolumn{2}{c}{}&\multicolumn{2}{c}{}&\multicolumn{2}{c}{}\\ \hline 
CHF$_3$$^k$ &   PBE    & 0&62 &19&63 &3074&5   & 1527&7$^d$ & 1139&2$^d$ & 1122&9 & 755&5 & 628&7$^d$ &\multicolumn{2}{c}{}&\multicolumn{2}{c}{}&\multicolumn{2}{c}{}\\ 
 &PBE+FCACP & 0&58 &19&29 &3153&9   & 1524&3$^d$ & 1153&5$^d$ & 1133&6 & 687&4 & 501&7$^d$ &\multicolumn{2}{c}{}&\multicolumn{2}{c}{}&\multicolumn{2}{c}{}\\ 
 &   PBE0   & 0&64 &15&25 &3158&3   & 1522&3$^d$ & 1212&4$^d$ & 1182&5 & 762&0 & 613&9$^d$ &\multicolumn{2}{c}{}&\multicolumn{2}{c}{}&\multicolumn{2}{c}{}\\ \hline \hline 
CCl$_2$F$_2$$^l$&PBE      & 0&21 &45&13 &1102&1 & 1058&0 & 867&0 & 723&7 & 499&6 & 492&7 & 460&8 & 440&0 & 275&9\\ 
&PBE+FCACP& 0&21 &44&69 &1119&3 & 1076&6 & 906&6 & 703&7 & 500&3 & 484&1 & 469&3 & 459&7 & 278&5\\ 
&PBE0     & 0&22 &35&44 &1189&4 & 1133&5 & 931&3 & 714&7 & 500&5 & 492&2 & 479&3 & 432&1 & 280&6\\ \hline 
CH$_2$Cl$_2$$^m$&PBE      & 0&62 &43&37 &3099&6 & 3021&7 &1401&4 &1226&7 &1139&0 & 876&3 & 720&9 & 693&6 & 263&5\\ 
&PBE+FCACP& 0&59 &42&99 &3169&5 & 3093&0 &1416&9 &1239&9 &1154&7 & 897&1 & 728&6 & 701&6 & 274&3\\ 
&PBE0     & 0&64 &34&33 &3190&3 & 3111&5 &1454&2 &1283&0 &1186&1 & 912&2 & 783&4 & 732&3 & 280&9\\ \hline 
CH$_2$F$_2$$^n$&   PBE    & 0&76 &18&50 &3073&0 & 2975&8 &1573&7 &1554&5 &1524&5 &1271&7 &1133&0 &1078&5 & 678&8\\ 
&PBE+FCACP & 0&72 &18&19 &3113&4 & 3043&2 &1495&7 &1419&7 &1240&1 &1192&3 &1080&7 &1043&0 & 567&1\\ 
&   PBE0   & 0&77 &14&42 &3159&8 & 3066&4 &1598&0 &1547&4 &1522&3 &1288&3 &1186&5 &1155&5 & 661&9\\ \hline \hline 
MAE & PBE             & 0&01 & 6&89 & 100&1 &   61&9 &   36&7&  32&3 &  17&7 &  30&6 &  44&8 &  41&2 &  13&0 \\ 
MAE &PBE+FCACP        & 0&03 & 6&56 &  33&3 &   31&7 &   37&6&  35&1 &  52&5 &  39&4 &  56&9 &  56&9 &  34&5 \\ \hline
\end{tabular}
\end{table*}

The transferability of the PBE+FCACP functional has been tested for 14 other molecules and properties.
Specifically, Table~\ref{tab:testresults} gives an overview for all the calculated dipole moments, 
polarizabilities, and harmonic frequencies for all test molecules.
As one would expect, the hybrid functional PBE0 yields consistently higher frequencies, 
lower polarizabilities, and lower dipole moments than the GGA functional PBE~\cite{ChemistsGuidetoDFT}.
For most molecules, the FCACP corrected GGA results indicate a clear and systematic shift towards the hybrid functional 
harmonic frequency numbers; not only for the molecules used for training (Table~\ref{tab:results}) 
but also for all the ``unseen'' test molecules: 
The mean absolute deviation from the PBE0 results for the test molecules reduces dramatically,
in case of the highest lying mode even from 100 to 33 cm$^{-1}$. 
As the frequency of the mode decreases, the deviation from PBE0 becomes smaller for both, PBE and PBE+FCACP. 
Eventually, for the lowest lying modes, the correction performs slightly worse than PBE, but is still within reason.
Encouragingly, we also note that PBE+FCACP rarely overestimates the PBE0 frequency, the largest overestimation
being $\sim$28 cm$^{-1}$ in the case of $\omega_8$ in CCl$_2$F$_2$.

As already alluded to above, the correction performs worst when Fluorine is present, albeit not always.  
For example, $\omega_4$ of CH$_2$F$_2$ is modeled worse by PBE+FCACP than by PBE alone, this is possibly also due to the fact that the PBE0 frequency is already smaller than the PBE frequency in this case. 
But also for CF$_4$ results for modes $\omega_2$, $\omega_3$, and $\omega_4$ suggest that there is still room
for improvement for the F correction.
 
For polarizabilities, the improvement is significantly less dramatic. 
As one would already expect from the optimization penalty results discussed in the previous section (Table~\ref{tab:params}),
the diagonal polarizability tensor elements change only in the single digit percentages.
The change, however, is consistently towards a reduction in polarizability, as desired.
The PBE+FCACP dipole moments worsen when compared to PBE, albeit ever so slightly. 

It remains to be seen if PBE+FCACP can also account for anharmonicities. 
In analogy to using multiple DCACP channels to generate the correct 1/$R^6$ dissociative behavior in the case of 
rare gas atoms~\cite{MultiCenterDFT_Tavernelli2009}, additional FCACP channels might be necessary. 
The PBE+FCACP forces on all atoms placed in the PBE0 geometry are typically rather small, but not negligible.
No imaginary frequencies have been found for any of the molecules investigated. 
A comprehensive assessment of these and further properties, however, is beyond the scope of this study, and will be the topic of future work.

\section{Conclusion}
A force correcting atom centered potential (FCACP) has been introduced for augmenting generalized gradient approximated (GGA) DFT calculations. 
FCACPs have been trained and tested using hybrid DFT reference calculations of various small molecules involving hydrogen, chlorine, fluorine and carbon atoms. 
The optimization of parameters has been accomplished by minimizing forces on atoms in hybrid DFT relaxed geometries, as well
as deviation of static polarizabilities from hybrid DFT values.
Numerical evidence suggests that not only hybrid DFT geometries can be achieved, but that also the corresponding harmonic frequencies
improve dramatically when using GGA+FCACP. 
The promising numerical results and the versatility of approach hold great promise  
that vibrational studies of condensed phase systems are
possible with hybrid DFT accuracy --- at the computational cost of GGA calculations.

In the future, it might be worthwhile to more systematically investigate the construction of better functionals using
series expansions of atom centered potentials on top of typical GGAs~\cite{AnnDivine}, 
and to assess the effect on multiple properties at once.
For example, one could combine FCACPs with DCACPs to calculate vibrational properties in molecular liquids or crystals, as well as phase diagrams.
It remains to be seen which of the many approximations to the exchange-correlation potentials are the most suited for being combined with atom centered potentials. 
Apart from PBE, the authors also considers the AM05 functional~\cite{AnnsSurfacefctlPRB2005} to be a potential candidate for such an effort. 
We do not know yet if such an GGA+FCACP+DCACP approach can account for all the many-body interactions present, 
recently found to be relevant even for van der Waals interactions~\cite{anatole-prb2008-2,anatole-jcp2010,mbd_PNAS2012}. 
Another critical issue is to more rigorously address the somewhat arbitrary choice of reference systems 
(molecules and geometries) and methods (other than hybrid DFT). 
Recent efforts using machine learning in chemical compound space might offer new strategies to remove
the resulting selection bias~\cite{RuppPRL2012}

\section{Acknowledgments}
This article is dedicated to Prof.~M.~Quack, the author's {\em Diplomvater} at ETH Z\"urich in 2001, 
and co-author of the resulting paper~\cite{anatole-pccp2007}.
The author is thankful for many technical discussions with 
P.~J.~Feibelman, A.~E.~Mattsson, and A.~G.~Taube at Sandia National Laboratories.
This research used resources of the Argonne Leadership Computing Facility at Argonne National Laboratory,
which is supported by the Office of Science of the U.S.~DOE under contract DE-AC02-06CH11357.
\bibliography{literatur}
\end{document}